\documentclass[10pt]{article}

\usepackage{amsmath,amssymb,mathtools}
\usepackage{xcolor}
\usepackage{natbib}
\usepackage[all]{xy}
\usepackage{tikz}
\usepackage{url}
\usepackage{tipa}
\usepackage{graphicx}
\usepackage{pgfplots}
\usepackage[
hidelinks,
colorlinks=true, 
linkcolor=blue,  
citecolor=black,
urlcolor=blue    
]{hyperref}
\usepackage{svg}
\usepackage[normalem]{ulem}
\usepackage{afterpage}
\usepackage[margin=2cm]{geometry}

\newcommand\Real{\mathbb{R}}
\DeclareMathOperator\proba{\mathsf{P}}

\newcommand\formantFrequency{F}
\newcommand\formant{\mathrm{F}}

\usepackage{authblk}

\begin{document}

\markboth{Tavakoli et al.}{Statistics in Phonetics}

\title{Statistics in Phonetics}

\author[1]{Shahin Tavakoli\footnote{address for correspondence: shahin.tavakoli@unige.ch}} 
\author[1]{Beatrice Matteo} 
\author[2]{Davide Pigoli} 
\author[3]{Eleanor Chodroff} 
\author[4]{John Coleman} 
\author[5]{Michele Gubian} 
\author[6]{Margaret E.\ L.\ Renwick} 
\author[7]{Morgan Sonderegger}
\affil[1]{Research Institute for Statistics and Information Sciences, Geneva School of Economics and Management, Universit\'e de Gen\`eve, 1205 Geneva, Switzerland}
\affil[2]{Department of Mathematics, King's College London}
\affil[3]{Department of Computational Linguistics, University of Zurich}
\affil[4]{Phonetics Laboratory, University of Oxford}
\affil[5]{Institute for Phonetics and Speech Processing (IPS), Ludwig Maximilian University of Munich}
\affil[6]{Department of Linguistics, University of Georgia \& Department of Cognitive Science, Johns Hopkins University}
\affil[7]{Department of Linguistics, McGill University}

\maketitle

\begin{abstract}
Phonetics is the scientific field concerned with the study of how speech is produced, heard and perceived. 
It abounds with data, such as acoustic speech recordings, neuroimaging data, or articulatory data.
In this paper, we provide 
an introduction to different areas of phonetics (acoustic phonetics, sociophonetics, speech perception, articulatory phonetics, speech inversion, sound change, and speech technology), 
an overview of the statistical methods for analyzing their data, 
and an introduction to the signal processing methods commonly applied to speech recordings.
A major transition in the statistical modeling of phonetic data has been the shift from fixed effects to random effects regression models, the modeling of curve data (for instance via GAMMs or FDA methods), and the use of Bayesian methods.
This shift has been driven in part by the increased focus on large speech corpora in phonetics, which has been driven by machine learning methods such as forced alignment.
        We conclude by identifying opportunities for future research.
\end{abstract}


\section{Introduction}

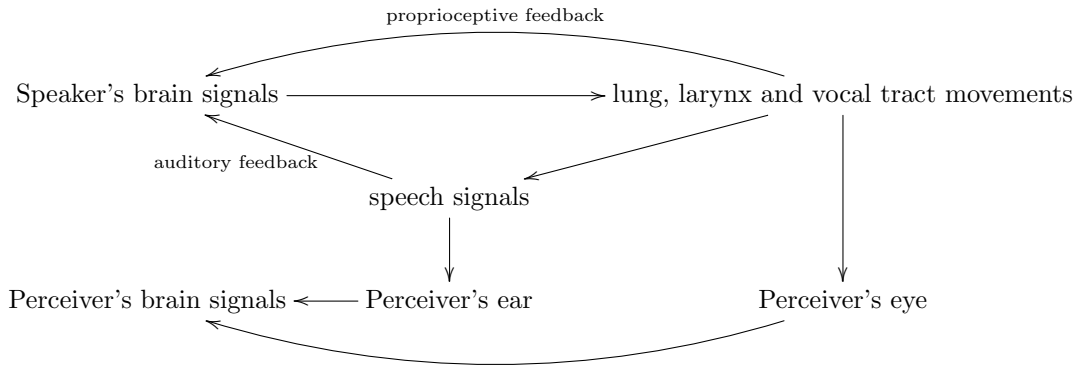
\begin{figure}[h]
\[
    \xymatrix{ \text{Speaker's brain signals} \ar[rr] && \text{lung, larynx and vocal tract movements} \ar[dd] \ar[dl] \ar@/_2pc/[ll]_{\text{proprioceptive feedback}}
                                                  \\ & \text{speech signals} \ar[d] \ar[lu]^{\text{auditory feedback \phantom{addf}}} & 
        \\  \text{Perceiver's brain signals} & \text{Perceiver's ear} \ar[l] & \text{Perceiver's eye} \ar@/^2pc/[ll]                               
}
\]
\vspace{.5cm}
\caption{The speech ``pipeline'': each arrow indicates one aspect of speech communication.} \label{fig:speech_pipeline}
\end{figure}

Speech, a fundamental form of human communication, is a very complex process that typically develops naturally from a young age.
 Phonetics is the primary scientific discipline that studies speech, including each aspect of speech communication shown in the diagram in \textbf{Figure~\ref{fig:speech_pipeline}}. 

\begin{figure}[h]
    \begin{center}
        \includegraphics[width=7in, trim=0 0 0 40, clip]{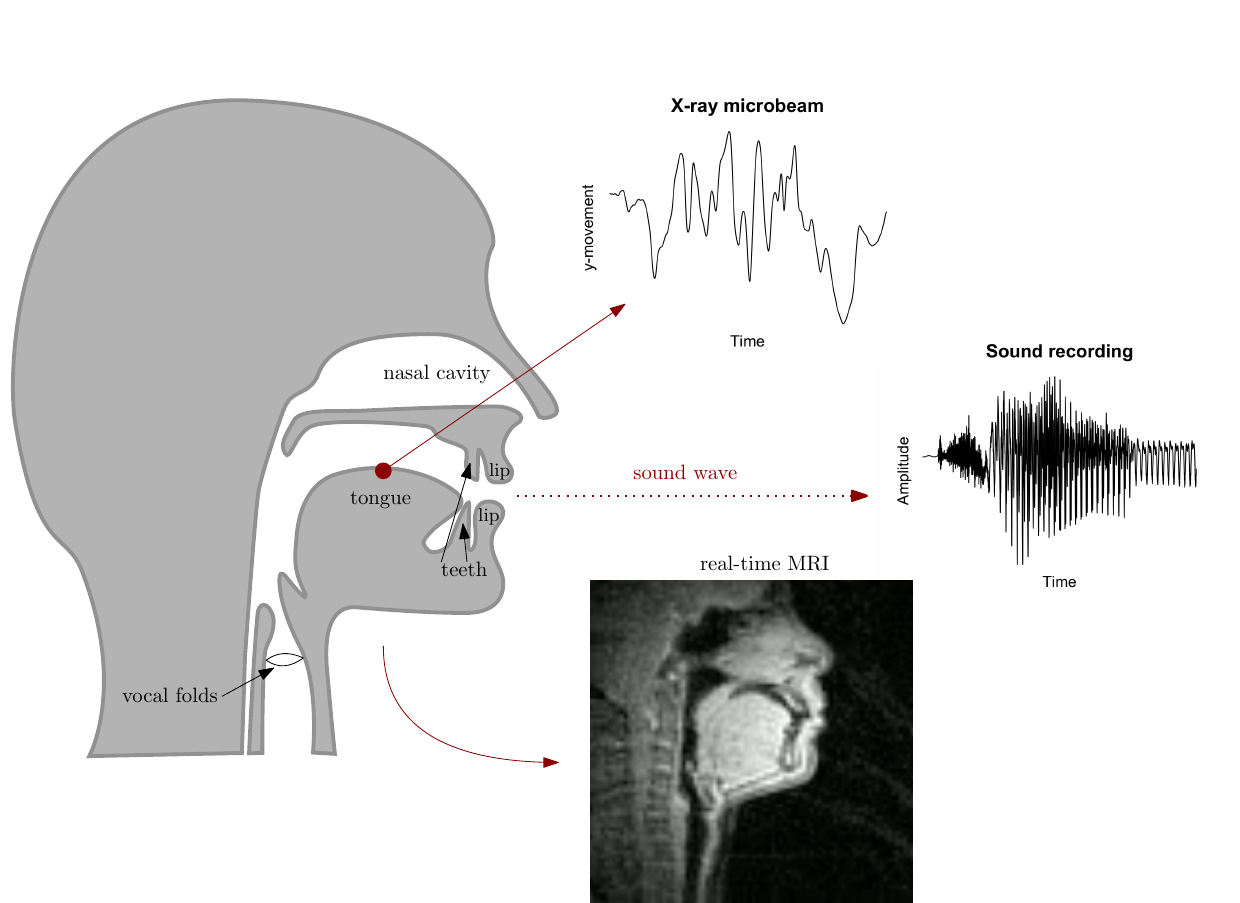} 
    \end{center}
    \caption{Some organs of speech, and examples of data that can be measured from them. The X-ray microbeam data is from \citet{westbury_x-ray_1994}. Real-time MRI data is from \citet{lim_multispeaker_2021}. We thank Wugapodes for the unlabeled version of the midsaggital head diagram, shared under CC0 1.0; \url{https://commons.wikimedia.org/wiki/File:Midsagittal_diagram_unlabeled.svg}.  }
    \label{fig:organs_speech}
\end{figure}

Phonetics can be thought of as a subfield of or closely allied field to linguistics, and also has close ties to  physiology, psychology, physics, machine learning, and sociology, among other domains.
As nicely expressed in \citet[][p.468]{brown_phonetics_2006}, 
\begin{quote}
    Phonetics attempts to provide answers to such questions as: What is the physical nature and
    structure of speech? How is speech produced and perceived? How can one best learn to pronounce
    the sounds of another language? How do children first learn the sounds of their mother tongue? How
    can one find the cause and the therapy for defects of speech and hearing? How and why do speech
    sounds vary---in different styles of speaking, in different phonetic contexts, over time, over
    geographical regions? How can one design optimal mechanical systems to code, transmit,
    synthesize, and recognize speech? 
\end{quote}
The acoustic signals produced in speech reflect the sounds of a language, which exist when then language is spoken or when audio recordings of speech are re-played. Sounds change historically: at a given time older speakers speak differently than younger generations, and the speech of each successive age group changes through history. Sounds are also personal: they reflect aspects of a speaker's social identity (such as age or gender), geographical dialect, and social group. Speech production requires complex and rapid coordinated movements of various speech organs (such as lungs, vocal folds, tongue, lips; see \textbf{Figure~\ref{fig:organs_speech}}), which the typical person takes for granted until they are made aware of them. The inventory of speech sounds, and the articulatory gestures used to produce them,  are not the same across languages or dialects. Speech production and speech perception are typically harder in a second language that is not learned in early childhood, though this is very much a matter of practice, motivation, and ability.

In this review, we focus on statistical methods  used in the analysis of speech data across several areas of phonetics, which we shall briefly introduce. 
{Given the broad scope of the paper, we cannot cover all of phonetics or all  statistical methods used in phonetics.}  Additional coverage is given by \citet{sonderegger_soskuthy_2024}, a review of statistical practice in phonetics written for phoneticians and speech scientists, whose approach is complementary to the current article.

We briefly describe the core concepts of speech production (\textbf{Section~\ref{sec:speech_production}}), then turn to representations of sound recordings (\textbf{Section~\ref{sec:frequency_representations}}). The core of the paper is \textbf{Section~\ref{sec:methods_and_open_problems}}, where we link different areas of phonetics to the statistical methods used to model specific types of phonetic data. \textbf{Section~\ref{sec:sociophonetics}} introduces acoustic phonetics and related topics (speech variation, sociophonetics), followed in \textbf{Section~\ref{sec:psycholinguistics}} by speech perception. We then turn to articulatory phonetics (\textbf{Section~\ref{sec:articulatory}}) and speech inversion (\textbf{Section~\ref{sec:speech_inversion}}). Sound change is presented in \textbf{Section~\ref{sec:sound_change}}, and 
\textbf{Section~\ref{sec:machine_learning}} treats the increasing role of machine learning in phonetics.  \textbf{Section~\ref{sec:conclusion}} concludes with some open questions.
\textbf{Section~\ref{sec:supplement}} provides lists of (mostly) freely available phonetic datasets, online books, and open-source software for analyzing phonetic data.

\section{Basics of speech production}
\label{sec:speech_production}

To produce most speech sounds, humans expel air from the lungs. Air first passes through the vocal folds, which can either remain open, be constricted, or close completely. When they remain open, the resulting sound is called {voiceless} or {unvoiced}. When they constrict, the vocal folds may vibrate, and the resulting sound is called {voiced}. The air, and the sound waves propagating through it, then pass through the {vocal tract}: first through the pharynx, then through either the oral cavity, or the nasal cavity, or both. The tongue modulates the sound as it passes through the oral cavity by making constrictions, where the tongue comes into close proximity or contact with parts of the oral cavity, thereby producing various sounds. The shape of the lips also affects the produced sound. The organs of speech, as well as examples of speech data, are depicted in \textbf{Figure~\ref{fig:organs_speech}}. 

The usual way to model the relationship between articulation, described above, and acoustics, the sound that is output at the lips, is 
the {source/filter model} of speech production \citep{fant_acoustic_1960,fant_relations_1980,stevens_acoustic_1998}. In this framework, voiced speech sounds start with the vocal folds vibrating, and the resulting sound wave passes through the vocal tract, which essentially acts as a filter, amplifying some frequencies and damping others.
A maximally simplified model for the filter is the {tube model} \citep{fant_acoustic_1960,johnson_keith_acoustic_2003}. It models the vocal tract as a series of connected tubes, either open or closed at their ends. From physics, we know that different lengths and diameters of the tubes give rise to different resonant frequencies, called {formants}, which are commonly used by phoneticians to characterize the produced speech sounds. 
While the tube model is an idealised model of speech production, it is a surprisingly good first approximation that explains how configurations of the vocal tract generate different speech sounds. Crucially, the tube model also motivates looking at sound recordings through the lens of Fourier analysis.

\section{Representations of the sound wave and derived features}
\label{sec:frequency_representations}

Speech generates air pressure variations which are perceived as sound. To record speech we can use a microphone, which has a diaphragm that vibrates under the action of sound waves, and this vibration is converted to an electrical current. This electrical current is then sampled at a given rate (typically 16 kHz\footnote{\emph{Hz and kHz}: {The hertz (Hz) is a unit of frequency equal to 1 cycle per second. For instance, $100$ Hz is $100$ cycles per second.  One kilohertz (kHz) is $1000$ Hz.}} for speech, or 44.1 kHz for the full spectrum of audible sound) and {quantized}\footnote{\emph{Quantization}: {The conversion of continuous values (e.g., $\tilde x \in \Real$) into an approximation $x$ from a finite set of values, typically encoded in bits. For instance, a signed $16$-bit quantization is a mapping $\Real \to \{-1,1\} \times \{0,1\}^{15}$}}, typically to signed 16-bit amplitude measurements.

\begin{figure}[p]
    \begin{center}
        \includegraphics[width=7in, trim=0 0 20 20, clip=TRUE]{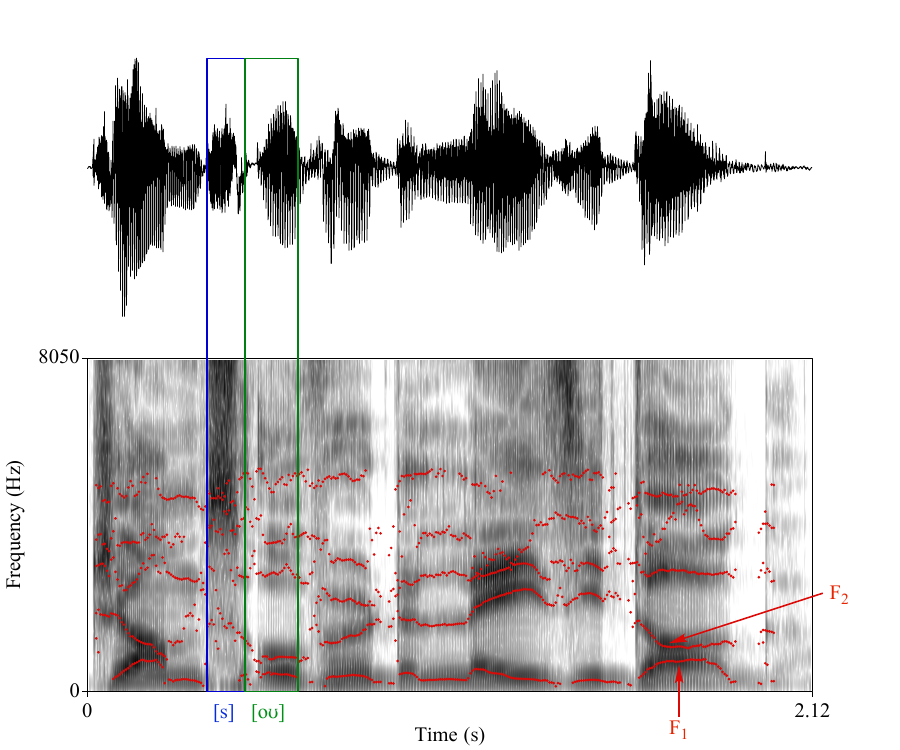}
    \end{center}
    \caption{Example of a speech recording $x[t]$ of the first author saying ``jumps over the lazy dog''. The top panel is an oscillogram (a plot of the amplitude $x[t]$ against the time $t$) and the bottom panel is its wide-band spectrogram with estimated formants (in red). Highlighted in blue is the segment corresponding to the fricative [s] (as in ``jump\textbf{s}'') and in green the segment corresponding to [o\textipa{U}] (as in ``\textbf{o}ver''). The first five formant frequencies are drawn in red on the spectrogram. Some estimated  formant frequencies match the spectral peaks in the wide-band spectrogram (mostly for the bottom ones, $\formantFrequency_1, \formantFrequency_2$) while higher formant frequencies ($\formantFrequency_3, \formantFrequency_4, \formantFrequency_5$) sometimes poorly estimate the spectral peaks. The figure was generated using the Praat software \citep{boersma_praat_2023}.}
\label{fig:example_sound_wave_spectrograms_formants}
\end{figure}

A (mono) speech recording can be viewed as a real time series $x[t]$, where $t$ is the time in seconds after the start of the recording, for instance $t = 0, 1/16000, 2/16000, \ldots$ if the sampling rate is $16$ kHz. {\textbf{Figure~\ref{fig:example_sound_wave_spectrograms_formants}}} (top panel) shows an example. In the language of time series, a  speech recording is not a stationary process \citep{shumway_time_2017}, but at the scale of about 10 ms ($10^{-2}$ seconds), most (but not all) segments are fairly stationary (stop consonants, such as [p]\footnote{It is customary to give phonetic transcriptions of speech sounds in square brackets.} as in ``\textbf{p}at'', may exhibit a short audible ``burst'' of about $5$ ms duration). For voiced sounds, the time series is (locally) well approximated as a sum of finitely many sine waves. For fricatives (e.g., [s] as in ``\textbf{s}ea''), the time series is closer to a sum of sine waves over a continuum of frequencies. Given the different characteristics of these sounds in the time domain, a frequency-domain approach to the analysis of speech recordings is often more informative. This approach is very important in the study of speech and language, as discussed from \textbf{Section~\ref{sec:sociophonetics}} onwards. 
A useful visual representation of speech recordings is a spectrogram, which is a time-frequency representation of the time series $x[t]$. 
To analyze the signal locally in time, say using an interval of 10 ms, the squared modulus of the fast Fourier transform \citep[FFT;][]{cooley_algorithm_1965}, or periodogram, is computed for $x[t]$. This gives the energy of each frequency component of $x[t]$, locally in time. The FFT is repeatedly computed by shifting the time window, resulting in a matrix, where each column corresponds to the periodogram of one time window, and each row corresponds to the energy of one frequency range over time. This matrix is plotted by mapping the energy values onto a color (or monochrome) scale, possibly after a logarithmic transformation (\textbf{Figure~\ref{fig:example_sound_wave_spectrograms_formants}}, bottom panel).  Another Fourier transform can be applied to the frequency range of the log-spectrogram, locally in time, resulting in the cepstrum \citep{bogert_quefrency_1963}. Cepstral analysis is useful as a method for separating the ``source'' and the ``filter'' (presented in \textbf{Section~\ref{sec:speech_production}}),
see for instance \citet{schafer_system_1970}.

The $y$-axis of the spectrogram, representing frequencies, may also be transformed non-linearly into scales that are more closely aligned to human perception, resulting in either the mel-, Bark- or auditory-spectrogram \citep{johnson_keith_acoustic_2003,gold_speech_2011}.
A discrete cosine transform is sometimes applied to the log mel-spectrogram (over the frequency bands), resulting in mel-frequency cepstral coefficients (MFCCs).  MFCCs are widely used in automatic speech recognition and synthesis (\textbf{Section~\ref{sec:machine_learning}}), though they have also been used in studies of speech variation (\textbf{Section~\ref{sec:sociophonetics}}) and sound change (\textbf{Section~\ref{sec:sound_change}}).
{\textbf{Figure~\ref{fig:continuum_speech_representations}} summarizes the computational links between these speech representations.

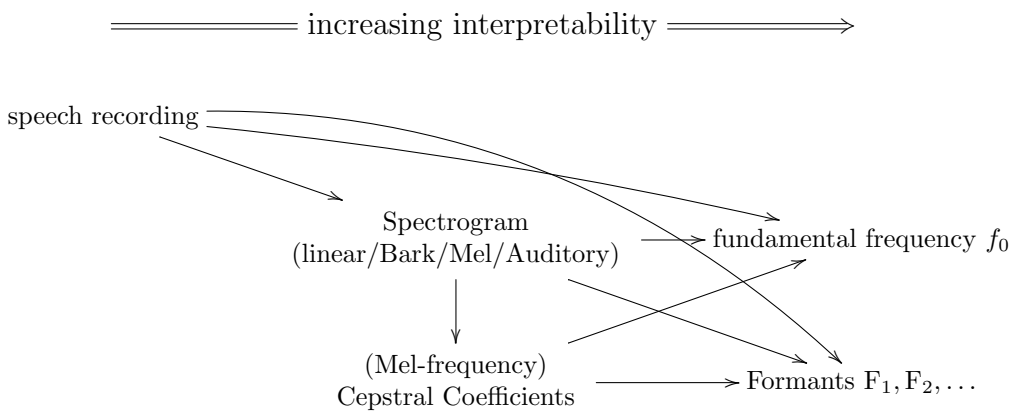
\begin{figure}[h]
\[
    \xymatrix{
        \ar@{=>}[rr]|{\text{\large~increasing interpretability~}} &&
            \\
       \text{speech recording} \ar@/^/[rrd] \ar[dr] \ar@/^3pc/[ddrr] && 
       \\  
                                                                                                                                                & {\begin{array}{c} \text{Spectrogram} \\ \text{(linear/Bark/Mel/Auditory)} \end{array}} \ar[r] \ar[dr] \ar[d] &   \text{fundamental frequency $f_0$}
                                                                \\ 
                                                                                                                                                & {\begin{array}{c} \text{(Mel-frequency)}\\  \text{Cepstral Coefficients} \end{array}} \ar[r] \ar[ru]   & \text{Formants $\formant_1, \formant_2,\ldots$}  
    }
\]
\caption{Common speech features/representations and the computational links between them.}
\label{fig:continuum_speech_representations}
\end{figure}

Perceptually important acoustic features of voiced speech include the fundamental frequency $f_0$, and the spectral peaks or {formants} $\formant_1, \formant_2, \ldots$.  Each formant has a centre frequency ($\formantFrequency_1, \formantFrequency_2, \ldots$), which is particularly important for distinguishing different kinds of vowels and consonants from one another.
Other acoustic features of speech (voiced or unvoiced) that are often analyzed include amplitude and duration, whether of sounds or other units corresponding to intervals of speech, such as syllables, words and phrases. Amplitude is typically measured by the root mean square (RMS) amplitude \citep[][Section~2.3.2]{johnson_keith_acoustic_2003}; it should not be confused with loudness, which is a perceptual quantity.

The fundamental frequency $f_0$, not to be confused with pitch (a perceptual property of how we hear and interpret the frequency of a sound), is the distance  between the peaks (harmonics) of the periodogram, typically measured in Hz (\textbf{Figure~\ref{fig:power_spectrum}}).   Methods for estimating $f_0$ include those given by \citet{boersma1993accurate} and \citet{talkin_robust_1995}.

On a given time window, the formant frequencies $\formantFrequency_1 < \formantFrequency_2 < \cdots$ are the frequencies of the peaks of the {spectral envelope}\footnote{\emph{Spectral envelope}: {The spectral envelope of a periodogram can be loosely described as ``the overall shape of the spectrum'' \citep[][p.97]{johnson_keith_acoustic_2003}  after abstracting away from the small bumps of the harmonics. It is often defined using the LPC spectrum, as shown in \textbf{Figure~\ref{fig:power_spectrum}}.}} of the periodogram \citep[][Section~4.2]{coleman_introducing_2005}, which can be visualized using a {wide-band spectrogram}\footnote{\emph{Wide-band spectrogram}: {A spectrogram obtained by using small time windows, of about 5 ms, giving low frequency resolution but high temporal resolution.}} (\textbf{Figure~\ref{fig:example_sound_wave_spectrograms_formants}}, bottom). 
To estimate the formant frequencies, a technique called {linear predictive coding} \citep[LPC;][]{atal_predictive_1978,schroeder_linear_1985,coleman_introducing_2005}
is typically used. In this approach,  an auto-regressive model of order $p$, or AR$(p)$ model \citep[see, e.g.,][Section~3.1]{shumway_time_2017}, is fitted to $x[t]$ via the {Burg algorithm}, which produces stable estimates.  The formant frequencies correspond to the peaks of the AR$(p)$ spectral density, also known as the LPC spectrum {(\textbf{Figure~\ref{fig:power_spectrum}}). The number of formants that can be detected is
approximately $(p-2)/2$, depending on the sampling rate used. With a sampling rate of 8 kHz, we can observe frequencies from 0 to 4 kHz, which will show about four formants, so $p=10$ LPC coefficients is appropriate. With sampling rate 16 kHz (8 kHz upper frequency limit), we can observe five or six formants, so $p=12$ or 14 LPC coefficients is appropriate.
Alongside each formant frequency $\formantFrequency_j$, its {bandwidth} $B_j$  \citep[a measure of the width of the peak;][pp.\ 153--162]{stevens_acoustic_1998} is also of interest, because it represents the range of acoustic frequencies excited by the formant and affects the perception of a speaker's voice timbre, among other things.

\begin{figure}[h]
    \begin{center}
        \includegraphics[width=7in, trim=0 0 20 0, clip=TRUE]{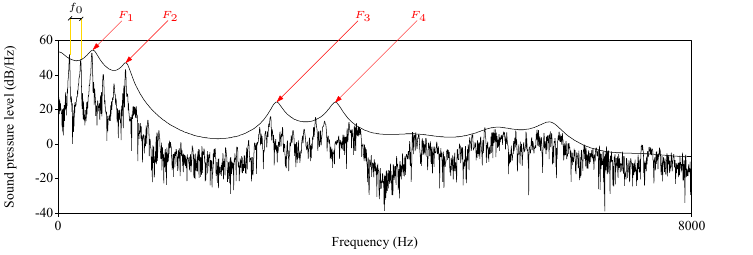}
    \end{center}
    \caption{Periodogram (jagged curve) and LPC spectrum (smooth curve) of a portion of one speaker's pronunciation of the vowel [o\textipa{U}] as in ``\textbf{o}ver'', with annotations for the fundamental frequency $f_0$ and formant frequencies. The LPC spectrum has been shifted vertically for ease of visualization.} 
    \label{fig:power_spectrum}
\end{figure}

\section{Statistical methods}
\label{sec:methods_and_open_problems}

\subsection{Acoustic phonetics}
\label{sec:sociophonetics}

Much research in phonetics falls under the category of acoustic phonetics \citep{johnson_keith_acoustic_2003}, which studies the acoustic realization of speech sounds, using the acoustic representations just discussed, such as spectrograms and formants.

\subsubsection{Sociophonetics}

In acoustic phonetics, a common problem is to quantify variation in the acoustic realizations of different consonants, vowels, patterns of pitch or other parameters of interest. For example, researchers may investigate  similar vowels across geographical space (i.e., dialects or languages), or between various social groups.
The latter is part of sociophonetics, a very active research area that emerged in the 21st century \citep{kendall_advancements_2023}. Sociophonetics studies how language is used to express social or context-dependent layers of meaning. 

{Sociophonetic analysis is inspired by sociolinguistic methods, which beginning in the 1980s treated spoken language variation in terms of categorical differences in pronunciation, using multivariate analysis programs like {Goldvarb and Varbrul} \citep{cedergren_variable_1974,tagliamonte_analyzing_2002}. 
    In modern sociophonetics, and in acoustic phonetics more generally, variation is often quantified with continuous phonetic measurements, as described in the following sections. 
}

\subsubsection{Analysis of scalar measurements}

Many studies focus on the analysis of one or a few scalar measurements. For instance, in the production of stop consonants (e.g.\ [p]), {voice onset time}\footnote{\emph{Voice Onset Time (VOT)}: {the interval of time between the release of a stop consonant (e.g. [p] or [b]) and the onset of voicing in the following sound.}} \citep[{VOT};][]{lisker_cross-language_1964,cho_voice_2019} is a key feature. Vowel pronunciations are often investigated using formant frequencies at mid-vowel \citep[][]{hagiwara1997dialect},
usually the first three formant frequencies $\formantFrequency_1, \formantFrequency_2, \formantFrequency_3$.
The first two formant frequencies $\formantFrequency_1, \formantFrequency_2$ have a nice interpretation in terms of articulation: $\formantFrequency_1$ is correlated with how much the tongue is lowered, and $\formantFrequency_2$ is correlated with the location of the tongue-palate constriction (i.e., tongue ``backness''), as shown in \textbf{Figure~\ref{fig:vowel_space}}.

\begin{figure}[h]
    \begin{minipage}{.6\textwidth}
    \includegraphics[width=\linewidth]{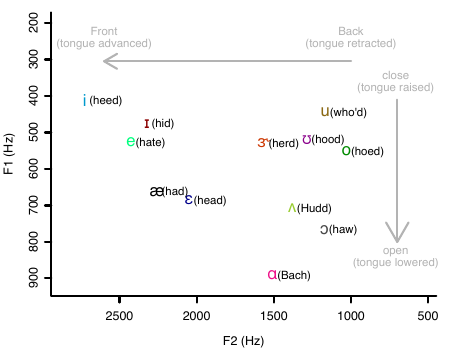}
\end{minipage}
    \begin{minipage}{.4\textwidth}
    \scalebox{-1}[1]{\includegraphics[width=\linewidth]{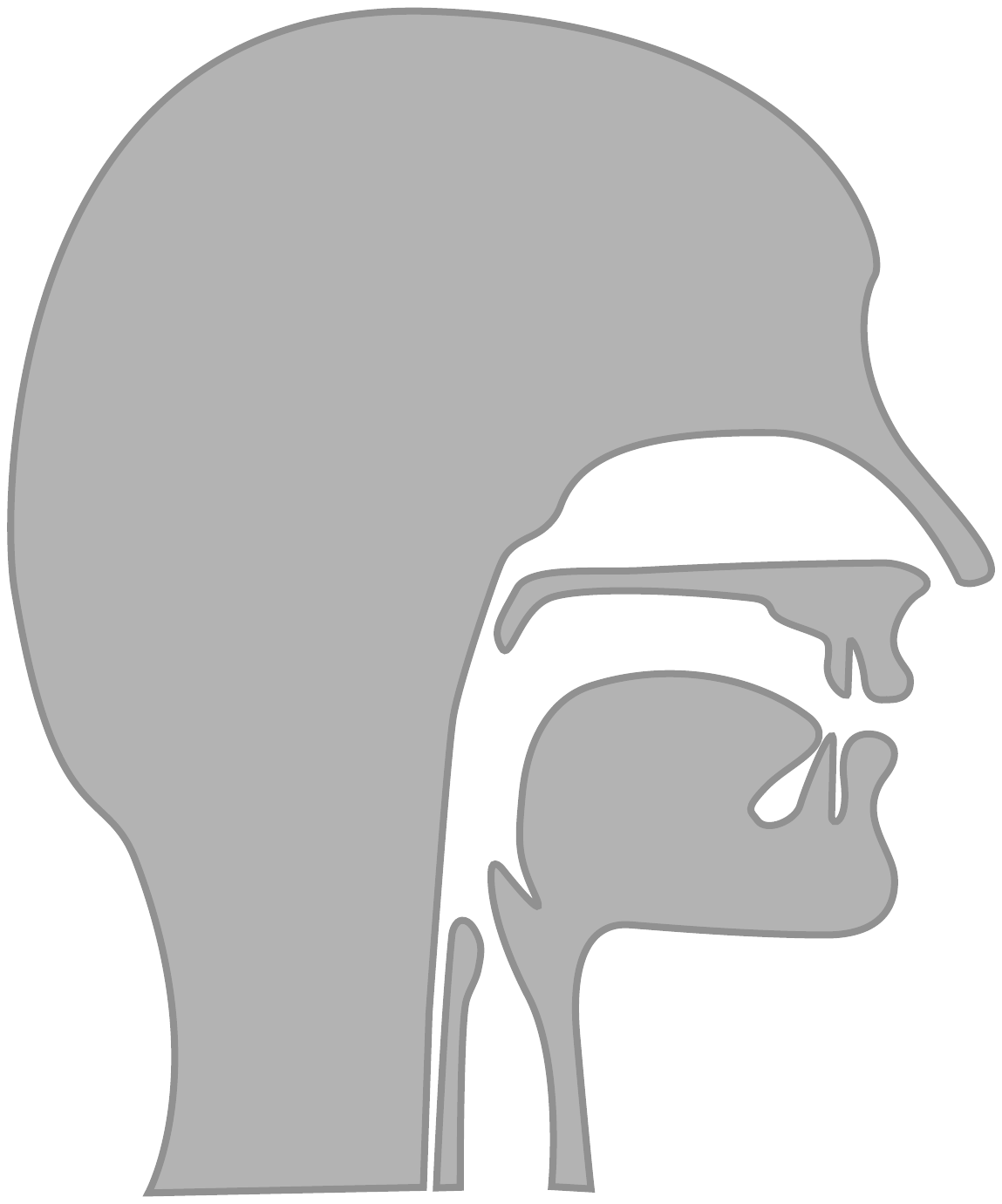}}
\end{minipage}
    \caption{ The $\formantFrequency_1, \formantFrequency_2$ vowel space (left sub-figure) of American English with average (raw) vowel formants over a set of 139 speakers. {Each vowel is written in the International Phonetics Alphabet,  with an American English pronunciation example in parentheses.}
The data used is from \citet{hillenbrand_acoustic_1995}, and available as the dataset \texttt{h95} in the R package \texttt{phonTools} \citep{phonTools}.
Note that the axis scales increase downwards and leftwards, in the opposite direction to mathematical convention; this is the tradition for such plots, so that the direction of changes in formant frequencies corresponds to the direction of tongue movements in a left-facing head (right sub-figure; we thank Wugapodes for the diagram on the right, shared under CC0 1.0; \url{https://commons.wikimedia.org/wiki/File:Midsagittal_diagram_unlabeled.svg}).
}
    \label{fig:vowel_space}
\end{figure}

Normalization of formant frequencies has been much discussed.
This is the problem of rendering formant frequencies from different speakers comparable to one another, allowing data to be pooled across people. Normalization is necessary because variation in speakers' physical characteristics (e.g., vocal tract length, mouth size) introduces variability in the formant frequencies that is usually irrelevant to the identity of the words, vowels and consonants being spoken, and does not disrupt accurate perception by listeners \citep{johnson2021speaker}. 
The goal of normalization is to remove such physiologically-derived differences while preserving both sociophonetic characteristics (such as dialectal variations, cross-language differences, and group differences) and distinctions between  different vowels. In addition to its use as a  tool to remove ``extraneous'' variation, normalization procedures are often of interest as models of the cognitive processes that allow us to recognise the same vowel when pronounced by different speakers \citep{rosner1994}.
There are numerous normalization methods, reviewed for instance by \citet{voeten_normalization_2022}. 
An example is Lobanov's (\citeyear{lobanov1971classification}) method, where the formant frequencies are $z$-scored per speaker, that is, the $j$-th formant frequency for the vowel $v$ is normalized as
\(
\formantFrequency^*_{j,v}= (\formantFrequency_{j,v} - \overline{\formantFrequency}_j)/s_j,
\)
where $\overline{F}_j$ is the average value of the formant frequency $\formantFrequency_j$ across all vowels within a speaker and $s_j$ is the standard deviation of the same sample. 
\textbf{Figure~\ref{fig:normalised_formants}} shows an example of raw and Lobanov-normalized plots of vowel formant frequencies.

Once normalization has been applied, subsequent statistical analysis aims to explore differences in acoustic parameters relevant to research questions: differences between groups of speakers or words, differences as a function of surrounding sounds, and so on.  The statistical models used initially were often simple visual comparisons, $t$-test and ANOVA models. However, as most phonetic experiments are based on measuring speech characteristics (such as VOT or formants) from a sample of speakers speaking a selection of words, possibly with replications, the field realized in the 2000s that ordinary linear models were often inadequate for these settings.
Since the goal of phonetic studies is to make inferences about the general population (either speakers of a certain language, or all humans), differences between speakers and words are better modeled with  participant and word random effects.  Phoneticians from the mid-2000s increasingly used repeated measures ANOVAs, then from around 2010 linear mixed-effects started to become the norm, due in part to the availability of the R package \texttt{lme4} \citep{bates_fitting_2015} and influential books and articles using R code \citep[e.g.,][]{baayen_analyzing_2008,baayen_mixed-effects_2008}; see \citet{sonderegger_soskuthy_2024} for a detailed account. In current practice, (generalized) linear mixed-effects modeling is ubiquitous in phonetic data analysis, and is the focus of current textbooks covering statistical analysis of phonetic data \citep[e.g.,][]{winter2019statistics,sonderegger_regression_2023}.

{The output of (generalized) linear mixed-effects modeling can be used as input to subsequent analyses. For instance, the speaker random intercepts can be used to assess the relative position of a person's speech compared to other speakers of the same age (to detect for instance leaders or laggers in sound change), and taken jointly over $\formantFrequency_1, \formantFrequency_2$ for various vowels they can be analysed via principal component analysis to detect systematic co-variation between vowels \citep{brand_systematic_2021}.}

\begin{figure}[h]
    \begin{center}
        \includegraphics[width=\linewidth]{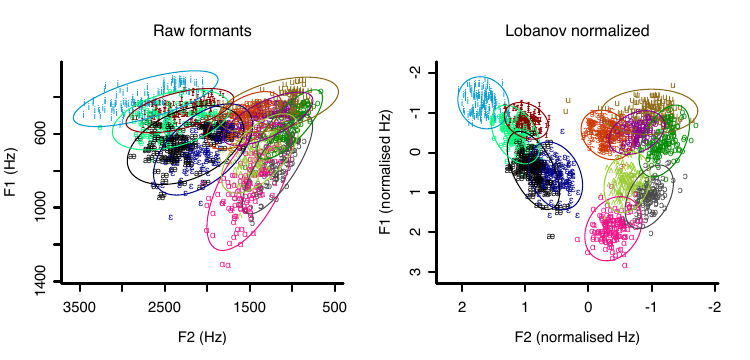}
    \end{center}
    \caption{Raw vowel space $(\formantFrequency_1,\formantFrequency_2)$ (left) and Lobanov-normalized vowel space (right) for the data in \citet{hillenbrand_acoustic_1995} available as the dataset \texttt{h95} in the R package \texttt{phonTools} \citep{phonTools}.  The superimposed ellipses are approximate $95\%$ confidence regions for each vowel. Note that the axes are reversed, as in \textbf{Figure~\ref{fig:vowel_space}}.}
    \label{fig:normalised_formants}
\end{figure}

Bayesian methods and philosophy \citep{gelman_bayesian_2013}  have become more popular in recent years \citep[facilitated by the R package \texttt{brms};][]{burkner_brms_2017}, a user-friendly interface to the probabilistic programming language Stan \citep{carpenter_stan_2016}. Bayesian models have allowed greater model flexibility and the integration of prior information about relevant variables to model speech data \citep[e.g.,][]{tanner2020toward,chodroff_uniformity_2022}, in contrast to previous methods that were essentially frequentist \citep{vasishth_bayesian_2018}. 
Bayesian models also have the convenient property of providing estimates of random effects when frequentist methods (such as those in \texttt{lme4})
fail to converge.
{However, to the best of our knowledge, the question of whether the Bayesian approach provides reliable estimates in such situations has not been clearly answered, either in theory or through simulation \citep[though see][]{eager2017mixed}}.
\citet[Section 3]{sonderegger_soskuthy_2024} provides a detailed review of Bayesian methods in phonetics.

\subsubsection{Analysis of trajectories}
\label{sec:analysis_of_trajectories}

For particular speech sounds, such as diphthongs (e.g.\ the vowel sound in ``loud''), the formant frequencies at mid-vowel alone do not capture the sound's dynamic nature, so it is necessary to model the entire time-course of parameters of interest. 
This led to the use and development of generalized additive mixed models \citep[GAMMs;][]{wood_generalized_2017,wieling_analyzing_2018},
functional data analysis \citep[FDA;][]{ramsay-sil-bookfda} approaches \citep{aston_linguistic_2010,tavakoli_spatial_2019,renwick_modeling_2020,koshy_exploring_2022}.

GAMMs can analyze dynamic phonetic data, such as formant frequency trajectories, articulatory movements, or $f_0$ trajectories, without simplifying or aggregating the data over time, and can flexibly capture the shape of the response curve 
while accounting for random effects of speakers and items (e.g., vowels). They can also be used to model a non-linear effect of a covariate on the response, such as the effect of age on formant frequencies \citep[e.g.][]{fruehwald2017generations}.
These models can be fitted in R  using the \texttt{mgcv} package \citep{wood2011fast}.
An example of application of a GAMM with non-parametric (fixed) effects and additive random intercepts and slopes for speaker, word uttered and dataset is given in \citet{renwick_boomer_2023}, modeling change over generational time in the pronunciation of vowels.
The \texttt{mgcv} implementation allows for random variation in curve shape at the speaker and item level, thanks to the link between penalisation parameters and random effects \citep[][]{wood_generalized_2017}, and a correlation structure in the errors can be included in the model. For an application to the comparison of tongue tip movement between native and non-native English speakers, see \citet{wieling_analyzing_2018}. These models have been used by \citet{chuang2021analyzing} to analyze response time in studies of high-level tones and ongoing {merger}\footnote{\emph{merger}: {Two sounds which were historically pronounced differently in a language come to be pronounced identically. For example, the spellings $\langle$ea$\rangle$ and $\langle$ee$\rangle$ in English word pairs like meat/meet, feat/feet reflect two distinct vowels in Middle English that are {merged} in modern English.  In southern US English dialects, the vowels [\textipa{E}]  and [\textipa{I}] are  merged before nasal consonants, so that words like ``pin'' and ``pen'' are pronounced identically.} \label{page:merger}} of two sets of sibilants in Taiwan Mandarin.
However, their complexity poses challenges in terms of inference and interpretation \citep{thul2021using}. Additionally, the computational cost can become high if a large number of basis functions must be used (especially with random variation in the curve shape).

Similar models  defined in the framework of functional mixed effects models \citep{guo2002functional} offer a larger variety of estimation approaches. For example, functional principal components can be used to reduce the number of basis functions in a data-driven way, so that standard estimation procedure for linear mixed-effects models can be used. This approach has been used to model fundamental frequency ($f_0$) in the tone language Qiang \citep{aston_linguistic_2010,evans_linguistic_2010}, and is available  through the R package \texttt{multifamm} \citep[][]{volkmann2023multivariate}, which allows the user to jointly model multiple response curves, such as the $f_0$ and formant frequency trajectories. Other models include, for instance, functional mixed effects models for irregularly sampled functional data \citep[][with the R package \texttt{sparseFLMM}]{pouplier2017mixed}.

More generally, a functional data analysis approach can help capture the time-dependent nature of phonetic features. For example, time-warping \citep[see, e.g.,][which discusses in particular time warping with the square-root velocity function and the Fisher--Rao metric, which do not seem to be well-known by phoneticians]{marron_functional_2015} can address the problem of time normalization for voice signals \citep{lucero_time_2000} such that time-variability can become part of the statistical analysis. \citet{koenig_speech_2008} explored differences in time variability and curve shape variability for oral airflow signals in fricative production between children and adults. Time-warping curves can also be included in the response of functional mixed effects models to treat individual variation in both phase and amplitude of the signal, such as in \citet{hadjipantelis2015unifying} for the analysis of $f_0$ trajectories in Mandarin Chinese. 
Other examples of functional data analysis techniques used in phonetics include modeling $f_0$ trajectories with orthogonal polynomials \citep{grabe_connecting_2007}, functional principal components for dimension reduction  to jointly explore formants and speech rate \citep{gubian_using_2015},
function-on-scalar regression for analyzing spectra of fricatives and stop releases \citep{Puggaardrode2022analyzing}}, nonparametric functional regression for spatial smoothing \citep{tavakoli_spatial_2019} and generalised linear models with functional predictors to capture differences in pronunciation \citep{koshy_exploring_2022}.

For a more in-depth review of methods for analyzing dynamic data in phonetics (e.g., formant trajectories),  see \citet[][Section 4]{sonderegger_soskuthy_2024}.

\subsection{Speech perception}
\label{sec:psycholinguistics}

In perception studies, experiments typically measure a subject's response to acoustic speech stimuli \citep{delattre_acoustic_1955,lisker_voicing_1970,pisoni_reaction_1974,kohler_categorical_1987}. These stimuli are often generated or manipulated artificially, for instance by varying VOT, formant frequencies or pitch.
Responses may be discrete motor responses (e.g., key presses representing listener judgements) or continuous signals (e.g., physiological measurements).
The data obtained in such experiments were initially modeled through linear models, followed by linear mixed models, similarly to the analysis of univariate pitch or formant frequency measurements (see \textbf{Section~\ref{sec:sociophonetics}}).
Even though early methods for sound manipulation generated odd-sounding or unrealistic sounds,
several methods now exist for generating good quality synthetic speech for use in perception experiments, such as Klatt synthesis \citep{klatt_analysis_1990}, LPC-based manipulation of pitch \citep[e.g.,][]{kohler_categorical_1987}, Tandem-STRAIGHT \citep{kawahara_tandem-straight_2008},  or MFCC synthesis \citep{erro_harmonics_2014,hudson_using_nodate}.

Other perceptual methodologies expose participants to auditory stimuli while simultaneously measuring brain activity, using for instance electroencephalography \citep[EEG;][]{ombao_handbook_2016} or magnetoencephalography \citep[][]{proudfoot_magnetoencephalography_2014}. The variable of interest is the brain signal change when an atypical auditory stimulus is presented. For each stimulus, event-related potentials (ERP\footnote{\emph{ERP}: {the EEG signal following a stimulus, see, e.g., \citet{kaan_eventrelated_2007}.}})  are recorded: these are often averaged at the subject level, and the ERPs of typical stimuli are compared to those of atypical stimuli, often using the mismatch negativity\footnote{\emph{mismatch negativity}: {the negative ERP peak in the range 100--250ms after stimuli, see, e.g., \citet{naatanen_early_1978}.}}. Working with EEG data involves a standard neuroimaging statistical toolbox \citep{ombao_handbook_2016}. When working with univariate summaries of the EEG signal, the methods used are similar to those described in {\textbf{Section~\ref{sec:sociophonetics}}}, with an emphasis on the use of random effects for modeling stimulus and subject effects. Working with the full multivariate EEG signal (the time series) requires the same level of statistical sophistication as analyzing data from functional magnetic resonance imaging studies \citep[fMRI;][]{ombao_handbook_2016}, in addition to the requirement of using random effects, such as multiple testing corrections for mass univariate approaches \citep{groppe_mass_2011}, and cluster detection \citep{frossard_cluster_2022}.

    Complementing brain activity measurement, experiments measuring the movements of the eyes---using the {visual world paradigm}\footnote{\emph{Visual world paradigm}: {a type of experiment in which participants hear an auditory stimulus (e.g., an uttered word) while looking at visual displays of the words and distractors.}} \citep{barr_analyzing_2008}---open a window into speech perception and processing as it happens. These experiments involve tracking eye movements of the subject following auditory stimuli. 
The eye movements define time series, which are often summarized by the proportion of gaze fixations to various areas of interest (such as visual depictions of words beginning with a certain sound). These are then analyzed by linear mixed-effects models or repeated measures ANOVAs and growth curve analysis \citep{mirman_statistical_2008}.
It is an ongoing debate among psycholinguists and phoneticians whether such analyses are sufficiently informative, or whether they should analyze the entire trajectories of eye movement (which in principle carry much more information, but are much harder to model), using for instance GAMMs \citep{nixon_temporal_2016}.

\subsection{Articulatory phonetics}
\label{sec:articulatory}

Articulatory phonetics, which studies how speech is produced through the movements of parts of the human vocal tract (see \textbf{Section~\ref{sec:speech_production}}), plays a key role in understanding the variability in speech production.
{Many studies in articulatory phonetics focus on the relationship between specific movements of the articulators (\textbf{Figure~\ref{fig:organs_speech}})---such as tongue shapes and lip rounding \citep[e.g.,][]{harshman_factor_1977}---and the acoustic output as seen in a spectrogram (\textbf{Figure~\ref{fig:example_sound_wave_spectrograms_formants}}, bottom panel).  For example, as illustrated in \textbf{Figure~\ref{fig:vowel_space}}, the vertical and horizontal position of the tongue body roughly determine the first and second formant frequencies. A longstanding question in articulatory phonetics is how good this first approximation is, and how other aspects of articulation (like lip rounding) affect formant frequencies. }

The available measurement technologies for tracking articulator movements constrain the modeling of articulatory gestures. 
Point-tracking methods, such as {X-ray microbeam}\footnote{\emph{X-ray microbeam}: {a technology in which gold pellets are attached to external and internal articulators on the midsagittal plane, and the movement of the pellets is automatically tracked using movable X-ray beams.}} \citep[][]{kiritani_tongue-pellet_1975,stone_three-dimensional_1990} and {electromagnetic articulography} \citep[{EMA}\footnote{\emph{EMA}: {an instrument in which coils of wire are placed on internal and external parts of the mouth and a variable electromagnetic field is created around the head of the patient, which can be used to estimate their position \citep{rebernik_review_2021}.} \label{note:EMA}};][]{pouplier2017mixed,wieling_analyzing_2018}, allow the measurement of particular points on the articulators over time. 
The advantage of such methods is their high temporal resolution, but the spatial resolution is limited (to about six points for EMA): interesting articulatory gestures could be missed if they do not appear at those points.
In contrast to this, medical imaging methods give measurements over a spatial continuum. Ultrasound is used especially to study tongue \citep{ultrasoundPINI2019162,davidson2006comparing,mielke2017ultrasound} and larynx \citep{moisik-larynx-ultrasound} gestures, but only soft tissue can be reconstructed with this imaging modality, which can be problematic for studying certain kinds of speech sounds, such as fricatives. Real-time magnetic resonance imaging is a recent measurement technique \citep[rtMRI;][]{toutios2016reviewMRI} which allows the measurement of mid-sagittal plane images of the vocal tract soft tissues in real time, but at a reasonably high temporal resolution \citep[e.g., a dataset of 83 frames-per-second temporal resolution and $84 \times 84$ pixels spatial resolution is provided in][]{lim_multispeaker_2021}.

As instrumentation for articulatory measurement is often quite invasive, foundational studies about various aspects of articulation tend to have very few speakers and therefore not much opportunity for comparative statistics, though some statistical tests used early on include ANOVA, Student’s $t$-test, the Wilcoxon signed-rank test, the Mann--Whitney $U$ test, the $\chi^2$ test, and the binomial test.
Later on, phoneticians became more aware of some of the problems of ANOVAs being used inappropriately (i.e., without checking for homogeneity of variance, normal errors, and independence of observations), and the methods used became more sophisticated, such as repeated measures ANOVA and MANOVAs.

Many recent statistical methods used in articulatory phonetics include functional data analysis \citep[][]{ramsay-sil-bookfda}. Indeed, even if the articulatory data is based on a point-tracking method, the time series nature of the measurements lends itself naturally to viewing the data as (multivariate) curves. In the case of rtMRI and ultrasound data, the data is of functional nature (in space) even for a fixed time point.  FDA was introduced to articulatory phonetics by \citet{ramsay1996lip}, who used functional PCA and functional ANOVAs to analyze lip motion. Given the subject and item levels present in most articulatory phonetic datasets, random effects need to be included in the model. These can be included through GAMMs \citep{tomaschek13_interspeech,wieling2016investigating} or  functional data approaches, such as in \citet{cederbaum_functional_2016}, which were compared in \citet{carignan2020analyzing}.
Our view is that these two approaches are very similar, particularly as the R package \texttt{mgcv} to fit GAMMs allows the inclusion of smooth random effects in the model \citep[see \texttt{factor.smooth} in the R package \texttt{mgcv},][pp.\ 326--7]{wood_generalized_2017}.
The difference lies in the basis functions used to model the curves, as \texttt{mgcv} mainly uses spline or similar bases, whereas functional data approaches allow the use of any functional basis.

\subsection{Speech inversion}
\label{sec:speech_inversion}

Speech inversion is the estimation of the  time series $y[t]$ of articulator shapes given a speech recording $x[t]$, that is, the acoustic-to-articulatory ``inverse map''  based on inversion of the articulatory-to-acoustic ``forward map'' of speech production, depicted in \textbf{Figure~\ref{fig:inverse_map}}.
\begin{figure}[h]
\[
    \xymatrix{ 
        \text{articulators} \ar@/^2pc/[rrr]^{\text{Speech Production (forward map)}} &&& \text{acoustic signal} \ar@/^2pc/[lll]^{\text{Speech inversion (inverse map)}}
}
\]
\caption{The speech production ``forward map'' and speech inversion ``inverse map''}
\label{fig:inverse_map}
\end{figure}
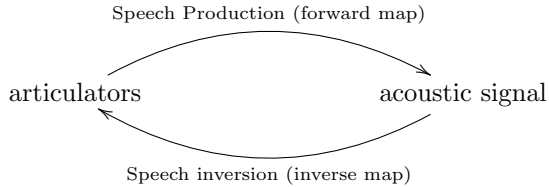

Speech inversion has a variety of applications---in clinical practice, phonetics, language learning and computer graphics---but it is challenging because it is a non-linear, ill-posed inverse problem (see \citealp{{kaipio_statistical_2005}} for  an introduction to statistical inverse problems). The acoustic-to-articulatory map is non-linear and one-to-many, so different vocal tract shapes may produce almost identical acoustic output \citep{Tukey1978InversionOA,mrayati_distinctive_1988,toda2004acoustic}.
Even in a lossless tube model of speech production, the vocal tract area function is not uniquely determined from the formant frequencies alone \citep{schroeder1967determination,mermelstein1967determination}. 
The one-to-many aspect of speech inversion comes mainly from the fact that different articulatory constrictions can create similar spectral filters \citep{mrayati_distinctive_1988}.
This happens in part because some articulatory constrictions produce {anti-resonances}, which dampen the energy at some frequencies \citep[for instance in nasal speech sounds,][Chapter~9]{johnson_keith_acoustic_2003}.
Broadly speaking, the speech inversion problem can be tackled either through acoustic models of the vocal tract, or through measurements of real (human-produced) articulatory/acoustic pairs.

Acoustic models of the vocal tract can describe the forward map, which is the  articulatory shape captured by cross-sectional area of the vocal tract \citep[see, e.g.,][]{beautemps_deriving_1995}. 
This map can be used to create an artificial training set of articulatory/acoustic pairs---an articulatory ``codebook''---to estimate the inverse mapping \citep{Tukey1978InversionOA,larar1988vector,schroeter_dynamic_1989,schroeter_evaluation_1990,rahim_acoustic_1991}. Alternatively, the forward mapping can be directly used in a hidden Markov model \citep[HMM; see, e.g.,][Chapter 9]{pml2Book}, with the hidden Markov chain being the sequence of discrete or continuous articulatory parameters, and the observed chain being the acoustic signal \citep{hiroya2004HMM}. 
However, acoustic models of the vocal tract are  far from an optimal approximation of the true forward mapping, because the vocal tract is irregular and composed of soft and hard tissue. This leads to energy losses due to yielding vocal tract walls, heat conduction and viscosity, together with radiation impedances when the sound wave leaves the lips and nose 
\citep{fant_vocal_1972,stevens_acoustic_1998}. 

Joint measurement of real (human-produced) articulation shapes and associated acoustic signals, as described in \textbf{Section~\ref{sec:articulatory}}, can be used as a training set for estimating the inverse mapping. This is not an easy task: rtMRI, for example, is a very loud 
procedure, which makes good audio recordings of speech difficult to obtain. However, the scanner noise may be mitigated using a combination of noise-cancelling microphones and post-processing of the recorded signals \citep{malinen_recording_2009}.

Nevertheless, Gaussian mixture models \citep{toda2004acoustic} or non-parametric smoothing \citep{ghosh_subject-independent_2011} have been applied to EMA data (see footnote on page~\pageref{note:EMA}).
Neural networks also constitute a natural approach to model the complex relationship between acoustic signal and articulatory configuration \citep{rahim_acoustic_1991}. 
For instance, \citet{richmond2006trajectory} used a mixture density network for speech inversion---essentially a mixture model whose parameters are predicted from the acoustic input through a multilayer perceptron \citep[][Section~13.2]{pml1Book}.
More recent approaches use deep learning methods to estimate EMA trajectories, XRMB trajectories, or even MRI images directly  \citep{seneviratne_multi-corpus_2019,siriwardena_acoustic--articulatory_2022}.

\subsection{Sound change}
\label{sec:sound_change}

Interesting statistical methods arise in historical phonetics, the study of how pronunciations change over time \citep{beddor_advancements_2023}, and comparative phonetics, the study of how the sounds of languages are related to one another.
Sound changes are often rooted in the exaggeration of inherent phonetic biases in articulation and perception. While they often arise due to such factors internal to a language, sound changes can also be caused by factors like social influences, contact between languages or long-term separation of related languages, sometimes as a result of major political changes (e.g., conquests) or geographical factors (e.g., migration). Sound changes can  have different effects on the phonological system of a language, such as creating, merging (see footnote on page~\pageref{page:merger}), or rearranging {phonemic contrasts}\footnote{\emph{Phonemic contrast}: {Two sounds are \emph{phonemes} that \emph{contrast} in a language if two words differing only in these sounds differ in meaning, and speakers agree that they sound different.  For example, we know /i/ and /o/ contrast in modern English because ``meet'' and ``moat'' are different words and any English speaker would say they are pronounced differently.  Phonemes are language-specific. For example, the distinction between aspirated \textipa{[t\super h]}, as in careful, standard pronunciations of ``water'', and the glottal stop \textipa{[P]}, as in colloquial, informal pronunciations of ``water'', does not constitute a phonemic contrast in English, while it does so in Hebrew, because there are word pairs in which those two consonants occur in distinct words, such as \textipa{[t\super hor]} ``queue'' vs. \textipa{[Por]} ``light''.} } that govern which words have distinct pronunciations. Sound changes are usually regular, meaning that they affect all words containing the relevant sounds or sound patterns, but  they can be irregular or sporadic, affecting only some words or forms (e.g., very high-frequency words or phrases, or words borrowed from other languages).
    An example of {borrowing} is the sound /\textipa{Z}/, which was adopted into English from French via words such as ``pleasure'' and ``rouge'' during a historical period of intense contact between the languages.
An example of regular sound change is the pronounciation of /t/\footnote{It is customary to give phonemic transcriptions in ``slash'' brackets.} in American English after stressed vowels (e.g., ``b\textbf{u}tter'', ``\textbf{a}tom'') as a ``tap'' closer to [d], where most British English dialects retain [t].

{Phonetic evidence has been central to the study of sound change since the 19th century. Work in sociolinguistics (the subfield of linguistics focusing on language in society, e.g., \citealp{labov_principles_2001,eckert2012three}),
has focused on how sound changes spread through a community and how they become integrated into a language's phonological system, emphasizing the interplay of social factors and speech perception/production.  Work in phonetics \citep[e.g.,][]{ohala_sound_1993,garrett_phonetic_2013,beddor_advancements_2023} has focused on the interaction of speech perception and production to explain why some sound changes occur much more frequently than others. An open question in sound change is the {actuation problem}: explaining why sound change occurs at some times but not others \citep{baker2011variability,yu2023actuation}, which has been the subject of computational modeling work \citep{niyogi_computational_2006,kirby_model_2013}.

Traditionally,  sound change has been studied using symbolic (i.e., alphabetic) representation of the normative way these sounds were pronounced in lists of cognate words \citep[e.g.,][]{jager2019computational}. A notable development has been the interpretation of sound changes in terms of vocal tract gestures \citep{browman1991gestural,carignan2021seed}, using the models discussed in \textbf{Section~\ref{sec:articulatory}}. 
However, such approaches have not yet been applied systematically to larger speech corpora.  

More recently, there has been a growing interest in modeling  sound change using speech recordings in historical phonetics, thus incorporating the variability of pronunciations within a language or dialect. This idea was first suggested by \citet{the_functional_phylogenies_group_phylogenetic_2012} by considering speech corpora from modern languages as data observed at the leaves of an evolutionary tree. This approach postulates that sound change can be described by a continuous process in some sound representation space (such as spectrograms or MFCCs), the relationships between languages represented by an evolutionary tree, with nodes representing clear separation of the direction of changes between two sub-branches of the tree. In this way, tree structure learnt from textual analysis or historical evidence can be incorporated into models with the hope of inferring past sound change, for example through Gaussian process regression \citep{rasmussen_gaussian_2005} where the kernel function is based on the tree structure \citep{the_functional_phylogenies_group_phylogenetic_2012}. Also, information from modern speech corpora can also be used to infer or verify hypothesised language tree structures \citep[e.g.,][]{shiers_gaussian_2017}. One important question for this approach is, what is the sound object that characterises the language pronunciation?   Average sound representations (such as spectrograms or MFCC, for example), as well as co-variability of these objects, have been considered \citep{pigoli_statistical_2018}. 
\citet{coleman_reconstructing_2015} used matrices of LPC parameters of whole words in order to model sound changes between them (e.g., Latin [tres] $\to$ Portuguese [trei\textipa{S}]); \citet{pigoli_statistical_2018} used FFT spectrograms of whole words as data objects, and \citet{hudson_using_nodate} use matrices of MFCC parameters as data objects representing word pronunciations.
However, the evolutionary models considered until now are too simple to be realistic and they suffer from the tension between the continuous representation of the sound change and the existence of apparently abrupt changes in the language \citep{pulleyblank2011abruptness}. 
Saltatory models based on Brownian motion have been suggested \citep{hudson_using_nodate} similar to those used in evolutionary biology, but this remains an underexplored topic.

Computational models have been proposed to reproduce sound change,  based on dynamical systems \citep{niyogi_computational_2006,sonderegger_combining_2010,kirby_model_2013} where 
the way parents transmit their pronounciation to children is modeled through speaker-listener exemplar-based probabilistic
models  \citep{bybee_phonology_2001,todd_word_2019}, and {agent-based models}\footnote{\emph{Agent-based models}: { computational models used to simulate and analyze how linguistic changes in pronunciation (sound changes) occur and spread within a population. In these models, individuals (agents) interact according to specified rules, and these interactions can lead to the adoption or evolution of new speech patterns. Agent-based models allow researchers to observe how certain factors, like social network structures or frequency of interaction, can influence the process of sound change over time. This approach provides a dynamic way to study language evolution, capturing complexities that other models might miss.  }} \citep{de_boer_origins_2001,Stevens2019,gubian_phonetic_2023}. The latter simulate sound change in a population, by modeling a population of speakers as different probability distribution-valued stochastic processes (agents) who are influenced by one another through complex interaction.

\subsection{Machine learning and speech technology in phonetics research}
\label{sec:machine_learning}

There have been continuous and mutual exchanges of knowledge between the fields of phonetics and speech technology.  Two representative speech technology tasks are automatic speech recognition (ASR) and text-to-speech synthesis (TTS). 
The goal of ASR is to transcribe spoken words as faithfully as possible. The goal of TTS is 
to produce natural-sounding speech from textual input. 

Until the early 2010s, speech technology was largely based on supervised learning. The problem of ASR was formalized as the maximization of $\proba(\text{word sequence} \mid \text{speech signal})$, which by Bayes' theorem can be decomposed as the maximization of the product of $\proba(\text{speech signal} \mid \text{word sequence})$---i.e., an acoustic model---and  $\proba(\text{word sequence})$---i.e., a language model \citep{Rabiner93}. The acoustic model was usually implemented as a hidden Markov model (HMM) over sequences of phonetic units, e.g., each HMM state mapping to a portion of a vowel or consonant, combined with a phonetic dictionary bridging between the (pre-processed) acoustic input and the sequence of written words. A similar, often richer, architecture was the base for TTS, where the role of HMMs was to generate sequences of speech parameters (fundamental frequency, vowel formants, etc.) based on the input text, which would then be turned into an acoustic signal by a vocoder \citep{zen2009statistical,erro_harmonics_2014}. All these systems required accurately-labelled speech corpora as training material. ASR required orthographically transcribed speech corpora associated with a phonetic dictionary as a minimum, while better training materials consist of manually-generated or manually-checked phonetic transcriptions (e.g.,  the TIMIT corpus, first released in 1988 and still in use: \citealp{garofolo1993timit}).

An important by-product of phonetic HMM-based ASR is the possibility of estimating the location of phonetic boundaries in the input speech signal, using the fact that the maximization of $\proba( \text{speech signal} \mid \text{word sequence})$ is carried out by applying the \citet{viterbi_error_1967} algorithm to the concatenation of the hidden states corresponding to the hypothesized phonetic sequence. The transitions between the hidden states obtained from the Viterbi alignment are used to estimate phonetic boundaries. A special case known as {forced alignment}\footnote{\emph{Forced alignment}: {automatic estimation of the timing of phonetic boundaries in an audio signal of which the phonetic transcription is known, e.g., the timing of the transitions between [k] and [\textipa{\ae}] and between [\textipa{\ae}] and [t] in the known sequence [k\textipa{\ae}t] (``cat''). It can be obtained as a by-product of HMM-based automatic speech recognition.}} consists in retrieving the timings of these boundaries for a known phonetic sequence, and then the Viterbi algorithm is run only once on the known hidden state sequence, rather than several times on many candidate sequences. Although of limited commercial interest, forced alignment represents a major contribution from speech technology to phonetic sciences \citep{yuan_speaker_2008,reddy_toward_2015,mcauliffe_montreal_2017}.
In particular, it facilitated the growth of corpus phonetics \citep{harrington_phonetic_2010,liberman2019corpus}, marking the transition from traditional phonetics, where controlled experiments are performed to collect limited amounts of speech data, to the analysis of observational data from speech corpora hundreds or thousands of hours in size (see \textbf{Section~\ref{sec:supplement}}).  

The last decade or so has been marked by the advent of deep neural networks (DNNs), which have revolutionized speech technology \citep{hinton2012deep}. Two major innovations characterize the current state-of-the-art of ASR, TTS and related tasks. One is the end-to-end approach, where system components designed on the basis of domain knowledge, like the phonetically informed acoustic models in ASRs described above, are replaced by a general-purpose computational architecture (DNN) trained using input-output associations, e.g., from sounds to words in the case of ASR. 
This has decreased the need for phonetically annotated corpora and phonetic dictionaries in the design of speech technology systems, particularly for high-resource languages (e.g., English, Mandarin), though significant challenges remain in ASR for low-resource languages (which are of significant interest for phonetics).  The second paradigm shift is the introduction of self-supervised representation learning \citep[SSL;][]{mohamed2022self}, where a DNN learns to predict its input or to reconstruct its randomly masked input. In the case of ASR, SSL is employed to learn context-rich numerical representations from untranscribed speech data sets of unprecedented size (decades). The pre-trained DNN is then fine-tuned, supervised by an orthographically transcribed (small) speech corpus.

\section{Conclusions and open problems in phonetics research}
\label{sec:conclusion}

Phonetics abounds with data, be it acoustic speech recordings (\textbf{Sections~\ref{sec:frequency_representations}--\ref{sec:sociophonetics}}), neuroimaging data (\textbf{Section~\ref{sec:psycholinguistics}}) or articulatory data  (\textbf{Section~\ref{sec:articulatory}}). Some landmark shifts in the statistical modeling of such data have been the introduction of random effects, the modeling of curve data (for instance via GAMMs or FDA methods), and machine learning methods such as forced alignment.
Some  open statistical modeling questions in phonetics are the following.

Many studies are based on the fundamental frequency $f_0$ and the formant frequencies. The accuracy of those measurements cannot be taken for granted, so they are manually checked in many studies. However the perspective taken is mostly a signal-processing one, and a more statistical approach is lacking: what  underlying parameter is being \emph{estimated} when computing $f_0$ and formant frequencies?
This presents a significant challenge, because unlike the relatively straightforward statistical problem of estimating (say) the expectation of a random variable $X$ from an independent sample $x_1,\ldots, x_n$, formulating a probabilistic model for a speech signal with specific values of $f_0$ and formant frequencies as its parameters is far more complex.

A limitation of the GAMM approach to phonetics data is that variable selection may be challenging. In phonetics, the need for better guidance on these procedures has been expressed \citep[e.g.,][]{soskuthy2021evaluating}. One possible solution is the use of penalisation (shrinkage) methods, such as the LASSO method \citep{hastie_elements_2009}, which have proved effective both for generalized additive models \citep{marra2011practical, bai2022spike} and for the selection and shrinkage of fixed \citep{schelldorfer2014glmmlasso} and random \citep{pan2014random} effects in generalised linear mixed effects models. To the best of our knowledge, the application of these methods to GAMMs has been limited to the case where only the fixed effects are modeled by smooth functions \citep{lai2012fixed}, though extending them to include smooth functions as random effects would be a useful development. {Note however that although most of the important acoustic variables of interest are smoothly changing, not all of them are: going beyond smooth functions would also be of interest.}

A big current challenge is to scale up speech models to take into account the growing amount of data now easily accessible online, such as audio recordings and audio-in-video of spontaneous, and continuous speech from many diverse speakers. Indeed, while forced alignment makes it possible to locate and extract hundreds of thousands of formant trajectories from vowel tokens in a large corpus, modeling such a quantity of multi-dimensional trajectories with GAMMs or other FDA techniques remains prohibitively expensive.

Extending current models to account for higher-level phonetic and linguistic aspects of speech---such as grammar, semantics, tone of voice (e.g., sarcasm, irony), discourse and conversational interaction---is of interest. This would shift the current focus on word pronunciations to how features of sound are used to encode higher-level linguistic and communicative structure.  Other challenges lie in modeling data that is quite easy to obtain in fairly large quantities---such as ultrasound tongue imaging---but difficult to process because it involves time series of very noisy images.

Modeling ``speech in the brain'', using technology that translates neural activity (measured for instance via electrode arrays) into speech is also a very active research area \citep[e.g.,][]{anumanchipalli_speech_2019}.

The recent developments of SSL for speech data have caught the attention of speech researchers, interested in discovering whether the representations learned from the speech signal (without supervision) bear any relation to concepts from phonetics, phonology, etc. Are phonetic and speaker information distinct in the SSL representation space? Do explicit representations of phones emerge? Is there any resemblance to the cognitive representations formed in the human brain? These hard questions come with a number of statistical challenges, as the analysis often starts with millions of large-dimensional vectors \citep{begus_ciwgan_2021,pasad2021layer,mohamed2022self,tom2022wav2vec}.

\section*{Acknowledgments}
We thank Stefano Coretta, Almond St\"ocker, Olivier Renaud, Dominique-Laurent Couturier, Josiane Riverin-Coutl\'ee and Meghan Clayards for helpful discussion and suggestions.
We also thank the editors and referee for comments that improved the quality of the paper.

\appendix

\section{Supplemental Material}
\label{sec:supplement}

The Supplemental Material include lists of (mostly) freely available phonetic datasets, online books, and open-source software for analyzing phonetic data.

\subsection{Online-only books}

\begin{description}
\item[Introduction to Bayesian data analysis for Cognitive Sciences] \citep{nicenboim_introduction_2023}
\item[Corpus phonetics tutorial] \citep{chodroff_corpus_2023}
\end{description}

\subsection{Datasets and software}
\label{sec:datasets_and_software}

\subsubsection{Speech corpora}

\begin{description}
    \item[Audio BNC] \citep{coleman_audio_2012}
    \item[Buckeye corpus] \url{https://buckeyecorpus.osu.edu/}  \citep{pitt_buckeye_2007}
    \item[Cross-linguistic voice onset time] \citep{chodroff_cross-linguistic_2019}
    \item[A Corpus for Large-Scale Phonetic Typology] \citep{salesky_corpus_2020}
    \item[SPADE] \citep{stuart-smith_spade_2022}
    \item[Doreco] a multilingual collection of recorded and transcribed fieldwork data from a diverse set of relatively understudied languages \url{https://doreco.huma-num.fr/}
    \item[VoxCommunis] acoustic models, pronunciation lexicons, and word- and phone-level alignments, derived from the publicly available Mozilla Common Voice Corpus \citep{ahn_voxcommunis_2022}. 
    \item[VoxAngeles] \url{https://github.com/pacscilab/voxangeles}, \citep{chodroff_phonetic_2024}
    \item[UCLA Phonetics Lab Archive] \url{http://archive.phonetics.ucla.edu}
    \item[SpeechBox] \url{https://speechbox.linguistics.northwestern.edu/#!/home}
    \item[Spontaneous speech corpora from Radboud University] \url{https://mirjamernestus.nl/Ernestus/Databases.php}
    \item[The Language Goldmine] provides links to linguistic databases and datasets \url{http://languagegoldmine.com/}
    \item[Corpus of Regional African American Language (CORAAL)] the first public corpus of African American Language data \url{https://oraal.uoregon.edu/coraal} \citep{KendallCorpusRegionalAfrican2023} 
    \item[Linguistic Atlas Project] outputs of research projects on American English, including spoken corpora available for download \url{https://www.lap.uga.edu/}
    \item[\href{http://lap3.libs.uga.edu/u/jstanley/vowelcharts/}{Gazetteer of Southern Vowels}] a web-based Shiny app for interacting with vowel data from the Digital Archive of Southern Speech \citep{StanleyGazetteerSouthernVowels2017}
    \item[\href{https://www.seas.ucla.edu/spapl/VTRFormants.html}{VTR Formants Database}] \citep{li_deng_database_2006}
\end{description}

A large number of other speech corpora are available, often for a small fee, from the \href{https://www.ldc.upenn.edu/}{Linguistic Data Consortium}.

\subsubsection{Acoustic-Articulatory datasets}
\mbox{}

\begin{description}
    \item[\href{https://www.cstr.ed.ac.uk/research/projects/artic/mocha.html}{MOCHA-TIMIT}] 
        a dataset is composed by the recordings of 460 sentences pronounced by 2 English speakers, a male and a female and the electromagnetic articulography trajectories for 8 articulators.  \citep{wrench1999mocha,mocha-timit-database-wrench-richmond}
    \item[\href{http://www.mngu0.org/}{mngu0}]  a corpus of articulatory data of different types (EMA, MRI, video, dental impressions, only partially available) with the corresponding speech recordings of 1300 sentences uttered by a single English speaker. \citep{richmond11_interspeech}
    \item[\href{http://www.cs.toronto.edu/~complingweb/data/TORGO/torgo.html}{TORGO}] a database \citep{rudzicz2012torgo} which collects recordings and EMA data from 8 subjects with dysarthria (5 male, 3 female) and 7 control subjects (4 male, 3 female). It contains recordings of sounds, short words and restricted and unrestricted sentences and EMA trajectories for 6 articulators. 
    \item[\href{https://yale.app.box.com/s/cfn8hj2puveo65fq54rp1ml2mk7moj3h/folder/30415804819}{Haskins Production Rate Comparison (HPRC)}] also referred to as the EMA-IEEE dataset \citep{tiede2017quantifying}, contains EMA data of 8 sensors trajectories and matching speech recordings of 8 English speakers (4 male and 4 female) articulating 720 sentences.
        \item[\href{https://sail.usc.edu/span/mri-timit/}{USC-TIMIT}] \citep{descr-TIMIT-narayan} a database that contains MRI and audio speech recordings from 10 speakers (5 male and 5 female) pronouncing the 460 TIMIT sentences. 
        \item[\href{https://figshare.com/articles/dataset/A_multispeaker_dataset_of_raw_and_reconstructed_speech_production_real-time_MRI_video_and_3D_volumetric_images/13725546/1}{A multispeaker dataset of raw [...] real-time MRI video and 3D volumetric images}] was released by \citet{lim_multispeaker_2021}. This database contains, 2D sagittal MRI videos with synchronized speech recordings, 3D volumetric vocal tract MRI and anatomical T2-weighted upper airway MRI.
        \item[\href{https://berkeley.app.box.com/v/xray-microbeam-database-data}{Wisconsin X-ray Microbeam Speech Production Database}] \citep{westbury1990x,westbury_x-ray_1994} 
        \item[\href{https://www.queensu.ca/psychology/speech-perception-and-production-lab/x-ray-database}{X-ray Film Database}] \citep{munhall1995x}. 
        \item[\href{https://epub.ub.uni-muenchen.de/47184/1/Articulatory_and_Acoustic_Characteristics_of_German_Fricative_Clusters.pdf}{Electropalatography data German fricative sequences}] Formant data Romanian vowel sequences
\end{description}

\subsubsection{Software}

A lot of resources can be found at the \href{https://lingmethodshub.github.io}{Linguistic Methods Hub} \citep{fruehwald_linguistics_2022}. Here are some specific softwares.

\paragraph{Speech annotation and analysis}

\begin{description}
    \item[\textsc{PRAAT}] Arguably the most used software in phonetics research \citep{boersma_praat_2023},
    \item[\textsc{ELAN}] Annotation software targeted to audio-visual corpora \url{https://archive.mpi.nl/tla/elan},
    \item[R package \texttt{emuR}] A speech database management and annotation in R \citep{winkelmann2017emu}
    \item[Speech Signal Processing Toolkit (SPTK)] \citep[][]{yoshimura_sptk4_2023}
    \item[R package \texttt{vowels}] for vowel normalisation \citep{KendallvowelsVowelManipulation2018}
    \item[R package \texttt{phonTools}] \citep{phonTools}
    \item[\href{http://darla.dartmouth.edu/}{DARLA: Dartmouth Linguistic Automation}] \citep{reddy_toward_2015} 
    \item[\href{https://phonetics.ucla.edu/voicesauce/}{VoiceSauce software for voice analysis}] \citep{Shuevoicesourcespeech2010} 
    \item[\href{https://github.com/santiagobarreda/FastTrack}{FastTrack formant tracking software}] \citep{BarredaFastTrackfast2021} 
\end{description}

\paragraph{Statistical analysis}    
\begin{description}
    \item[R package \texttt{lme4}] \citep{bates_fitting_2015}
    \item[R package \texttt{mgcv}] \citep{WoodmgcvMixedGAM2017}
    \item[R package \texttt{itsadug}] \citep{RijitsadugInterpretingTime2017}
    \item[R package  \texttt{fda}] \citep{ramsay-sil-bookfda}
\end{description}

\paragraph{Forced alignment}   
\begin{description}
    \item[\href{https://clarin.phonetik.uni-muenchen.de/BASWebServices/interface/WebMAUSBasic}{WebMAUS automatic speech recognition and forced alignment}] \citep{kisler_bas_2016} 
    \item[\href{https://montreal-forced-aligner.readthedocs.io/en/latest/}{Montreal Forced Aligner}] \citep{mcauliffe_montreal_2017}
\end{description}


\bibliographystyle{my-agsm}
\bibliography{biblio_paper.bib,shortened_biblio.bib}

\end{document}